\documentclass[aps,pra,twocolumn,showpacs,superscriptaddress,groupedaddress]{revtex4-1}

\usepackage{graphicx,amsmath,amssymb,amsthm,braket,mathtools,xcolor}

\newcommand*{\THETITLE}{Estimation of squeezing in a nonlinear quadrature of a mechanical oscillator}

\begin{document}

\title{\THETITLE}

\author{Darren W. Moore}
\email{darren.moore@upol.cz}
\affiliation{Department of Optics, Palack\'{y} University, 17. listopadu 1192/12, 771 46 Olomouc, Czech Republic}
\author{Andrey A. Rakhubovsky}
\affiliation{Department of Optics, Palack\'{y} University, 17. listopadu 1192/12, 771 46 Olomouc, Czech Republic}
\author{Radim Filip}
\affiliation{Department of Optics, Palack\'{y} University, 17. listopadu 1192/12, 771 46 Olomouc, Czech Republic}

\begin{abstract}
Processing quantum information on continuous variables requires a highly nonlinear element in order to attain universality. Noise reduction in processing such quantum information involves the use of a nonlinear phase state as a non-Gaussian ancilla. A necessary condition for a nonlinear phase state to implement a nonlinear phase gate is that noise in a selected nonlinear quadrature should decrease below the level of classical states. A reduction of the variance in this nonlinear quadrature below the ground state of the ancilla, a type of nonlinear squeezing, is the resource embedded in these non-Gaussian states and a figure of merit for nonlinear quantum processes. Quantum optomechanics with levitating nanoparticles trapped in nonlinear optical potentials is a promising candidate to achieve such resources in a flexible way.  We provide a scheme for reconstructing this figure of merit, which we call nonlinear squeezing, in standard linear quantum optomechanics, analysing the effects of mechanical decoherence processes on the reconstruction and show that all mechanical states which exhibit reduced noise in this nonlinear quadrature are nonclassical.
\end{abstract}

\maketitle

\section{Introduction}

Quantum states of oscillators which, in principle, have an arbitrarily large information capacity are attractive platforms for quantum technology. Quantum information processing with continuous variables (CV) is therefore a fast growing topic of research, at first conceptualised in the modes of the electromagnetic field~\cite{lloyd2003quantum}, then further finding a foothold in the vibrational modes of trapped ions~\cite{gulde2003implementation} and still further in the centre of mass motion of a macroscopic oscillator coupled to radiation pressure~\cite{schmidt2012optomechanical}. This last, the field of optomechanics, embodies a large scope of research into quantum technologies with proposals for sensing~\cite{doolin2014multidimensional,purdy2017quantum,armata2017quantum}, quantum communication~\cite{stannigel2010optomechanical,vostrosablin2016pulsed,ottaviani2017multipartite}, quantum computation (particularly the measurement based model)~\cite{houhou2015generation,moore2017arbitrary}, and tests of quantum gravity~\cite{pikovski2012probing} and foundations~\cite{marinkovic2018optomechanical,carlesso2018noninterferometric}. Linearised quantum optomechanics is very well established, both theoretically and experimentally with various platforms having demonstrated ground state cooling~\cite{teufel2011sideband,chan2011laser} and the preparation of squeezed states in the mechanical portion of the system~\cite{wollman2015quantum,pirkkalainen2015squeezing,lecocq2015quantum}. The time is ripe then, to begin looking for ways to add nonlinear mechanical effects to these achievements. 

At their most elevated station nonlinear elements are a necessary component of universal quantum computation with CV~\cite{lloyd2003quantum,menicucci2006universal}. However even before one goes so far, nonlinearity can be a useful resource for a variety of nascent quantum applications. To be more specific, there are several no-go theorems for Gaussian quantum information processing~\cite{magnin2010strong,jabbour2015interconversion} including entanglement distillation~\cite{fiurasek2002gaussian,eisert2002distilling,giedke2002characterization} and error correction~\cite{niset2009nogo}. Alongside these are some known applications for non-Gaussian resources~\cite{zhuang2018resource,albarelli2018resource} such as estimation~\cite{adesso2009optimal}, cloning~\cite{braunstein2001otimal}, teleportation~\cite{olivares2003teleportation} and Bell inequality testing \cite{paternostro2009violations}. In optomechanics, one has the advantage that the dynamics between optics and mechanics is intrinsically nonlinear. While the intrinsic cubic radiation pressure coupling is usually too weak to be considered useful for quantum technology (however see Refs \cite{brawley2016nonlinear,leijssen2017nonlinear}), the standard linearisation is an approximation that has the potential to be extended to a nonlinear regime involving the square of the mechanical position. Glimmers of such a future are visible in current electromechanics experiments~\cite{pirkkalainen2015squeezing,wollman2015quantum}, and proposals for taking advantage of the rich dynamics this extension entails already exist~\cite{tan2013generation, brunelli2018unconditional,brunelli2018linear}. Moreover, optomechanical couplings involving only the square of the mechanical position have been explored in multiple experiments~\cite{thompson2008strong,karuza2012tunable,paraiso2015positionsquared,kalaee2016design}. A versatile platform for nonlinear optomechanics is levitated optomechanics which, with recent developments in experimental techniques (i.e. coherent scattering), provides the opportunity to cool levitated nanoparticles to the ground state~\cite{windey2019cavity,delic2019cavity}, while also providing the opportunity to employ nonlinear potentials as external drivings for the oscillator~\cite{gieseler2013thermal,kiesel2014cavity,fonseca2016nonlinear,rashid2016experimental,siler2017thermally,ricci2017optically,setter2019characterization}. Various proposals for the generation of nonlinear and nonclassical states are extant in the literature~\cite{tan2013generation,hoff2016measurement,latmiral2018deterministic,li2018generation,lee2018dissipation,clarke2019growing}. Progress in this field is moving very fast and therefore it is important to analyse proof-of-principle possibilities to estimate what we refer to as the nonlinear squeezing in experiments. Therefore we consider in the abstract oscillators that have achieved nonlinearity directly through the coupling (as in membrane-in-the-middle setups), are intrinsically nonlinear (anharmonicity) or are prepared in states only achievable through the application of a nonlinear potential.

The preparation of a quantum cubic phase state, as an example of a nonlinear phase state, of a mechanical oscillator has been proposed using a variety of methods, including dissipative engineering~\cite{houhou2018unconditional}, stroboscopic pulses~\cite{rakhubovsky2019stroboscopic} and externally applied nonlinear potentials~\cite{siler2018diffusing}. Dissipative methods use a linear-and-quadratic interaction to create a nonlinear coupling to a cold reservoir, unconditionally cooling the mechanical state to a cubic phase state. The stroboscopic method involves short pulses of nonlinear potentials that can be applied to levitating nanoparticles~\cite{konopik2018nonequilibrium}, electrically actuated disk resonators~\cite{sridaran2011electrostatic} or mirrors coupled to optical springs~\cite{corbitt2007alloptical}. Levitated systems have the option to take advantage of, for example, Duffing nonlinearities~\cite{ricci2017optically} possibly modified by an external electric field, in order to attempt to mimic cubic nonlinearities. More broadly, the potential for the nanoparticle is determined by the optical intensity of the trapping field, something that appears to be under strong experimental control~\cite{romeroisart2011optically}. Outside the strict use of continuous variables, coupling to discrete variable elements can help to provide nonlinear behaviour~\cite{abdi2016dissipative,park2018deterministic}. These methods could be extended to also generate other higher order quantum nonlinearities. The cubic phase state is the simplest example in the class of nonlinear phase states~\cite{marek2018general} necessary to reduce noise in nonlinear circuits. The preparation of cubic phase states can be quite challenging and their verification resource intensive as can be intuitively seen from the complexity of their representation in phase space. Additionally, states with such a complex representation in phase space and detailed nonclassical features are usually easily smeared out by noise processes.

In this article therefore, in order to evaluate the nonlinearity of the prepared state, we propose a method to directly estimate the squeezing of any nonlinear quadrature, taking for clarity the simplest case of mechanical oscillators prepared in a cubic phase state. Normally squeezing means a reduction of the fluctuations in the variance of a variable below the value corresponding to the ground state. In linear oscillators, if there is squeezing it is always present in a variable which is a linear combination of position and momentum. At the same time squeezed states are nonclassical from the point of view of classical coherence theory~\cite{glauber1963quantum}. In nonlinear oscillators, squeezing (i.e. fluctuations below the level set by the ground state) can be found in a nonlinear combination of position and momentum even if not present in the linear case. We refer to this new object, first introduced in Ref.~\cite{miyata2016implementation} for light, as nonlinear squeezing to make clear the distinction from the linear squeezing in previous experiments described by linearized quantum dynamics in the Heisenberg picture~\cite{chan2011laser,wollman2015quantum,pirkkalainen2015squeezing,lecocq2015quantum}. The salient point is that nonlinear squeezing is necessary for the application of the cubic phase gate~\cite{miyata2016implementation}. Moreover, pure states which are nonlinearly squeezed are inherently nonclassical even for weak nonlinearities. We demonstrate that the extent to which states possess the property of nonlinear squeezing (and therefore nonclassicality) can be reconstructed via homodyne detection of the output cavity field without full tomography of the mechanical state and that this reconstruction is robust against noise for a wide range of experimental parameters. It follows that the reconstruction simultaneously allows direct identification of the nonclassicality of the mechanical cubic phase state.  
\begin{figure}
\includegraphics[width=\columnwidth]{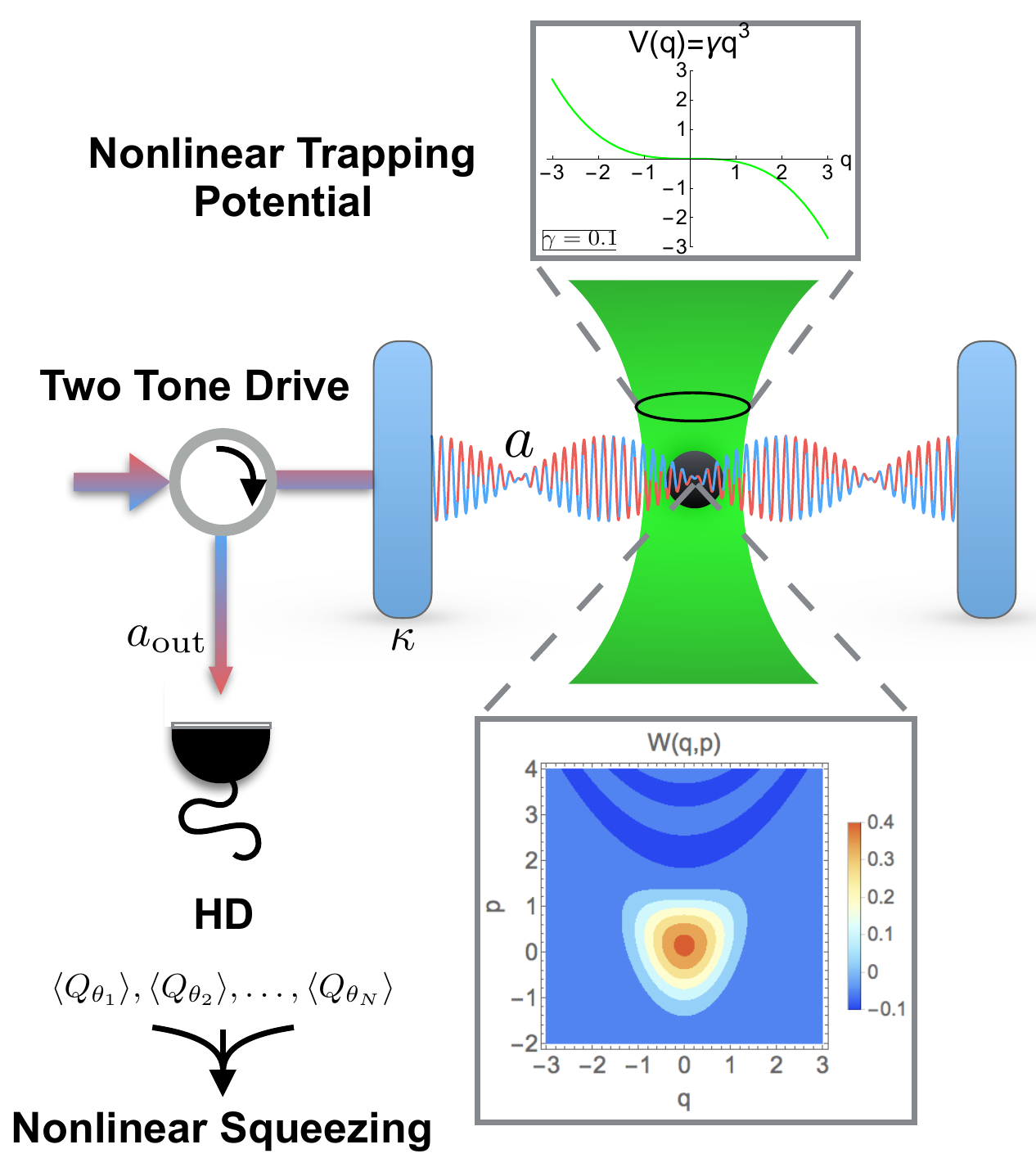}
\caption{A sketch of the proposed setup for a levitating nanoparticle motivated by the experiments in Refs~\cite{fonseca2016nonlinear,ricci2017optically}. The particle is first prepared in a cubic phase state, for example through the application of a Nonlinear Trapping Potential~\cite{siler2017thermally}. Afterwards, invoking linear optomechanics and applying a Two Tone Drive on the red and blue sidebands simultaneously provides the mechanism for readout of the nonlinear squeezing by inducing a QND interaction between the cavity momentum and a single mechanical quadrature. Subsequent to the interaction the output cavity field (dissipating at a rate $\kappa$) is measured via homodyne detection (HD). From the measured moments of quadratures the nonlinear squeezing and nonclassicality of mechanical states can be estimated.}\label{OMsketch}
\end{figure}

\section{Results}

\subsection{Nonlinear Squeezing and Nonclassicality}\label{NLS}

Squeezed states with the variance $\text{Var}(p_\text{LQ})$ of the quadrature $p_\text{LQ}=p-\lambda q$, where $\lambda\in \mathbb{R}$ is a parameter and $q,p$ are canonical position and momentum operators, suppressed below the ground state are a useful resource to implement quadratic nonlinearities in circuits using Gaussian measurements~\cite{filip2005measurement,miwa2014exploring,miyata2014experimental}. Advantageously, the purity of these squeezed states is irrelevant and only the variance matters. All such squeezed states are nonclassical~\cite{glauber1963quantum}, therefore their coherence effects go beyond classical states of oscillators. To deterministically implement higher than quadratic nonlinearities for quantum circuits, the measurement-induced strategy requires new ancillas beyond squeezed states. The principal example of such an implementation is the nonlinear cubic phase gate required for universal quantum computing with continuous variables~\cite{gottesman2001encoding}. The cubic phase state $e^{i\gamma_G q^3}\ket{p=0}$, in which $\ket{p=0}$ denotes a zero momentum eigenstate, may act as a resource for implementing the cubic phase gate with nonlinearity strength $\gamma_G$ as in the protocol of adaptive non-Gaussian measurements~\cite{miyata2016implementation}. More generally, implementing a nonlinear phase gate of order $n$ requires noise reduction in the nonlinear quadrature $p-n\lambda q^{n-1}$~\cite{marek2018general}. Fortunately, the nonlinear measurement strategy implies that the nonlinear quadrature $p_\text{NLQ}=e^{i\gamma_G q^n}pe^{-i\gamma_G q^n}=p-n\gamma_G q^{n-1}$ is the {\it only} relevant feature of the cubic phase state that allows the protocol to be carried out. Ideally, it is required that $p_\text{NLQ}$ vanish for the resource state consumed during the measurement process. If this is the case the nonlinear phase gate is applied in the output of the strategy. In our example, the unstable cubic potential depicted in Fig.~\ref{OMsketch} gives rise to a complex non-Gaussian Wigner function, also present in the figure, of the motional state with negative values indicating a highly nonclassical nature. These negative values of the Wigner function are sensitive to loss and noise in state preparation and estimation. Importantly, neither the purity nor any variable other than $p_\text{NLQ}$ is relevant for implementing the gate. 

In practice the perfect cubic phase state is inaccessible, being unphysical, and approximations to the ideal case must be used. This unphysical character manifests itself with two aspects: the infinite squeezing of the momentum eigenstate $\ket{p=0}$ and the unbounded character of the cubic potential. These properties are approximated using finite squeezing and an appropriate bounded version of the unbounded cubic potential respectively. An alternative solution is to search for other states sharing the relevant properties~\cite{miyata2016implementation}. Such states also form a resource for implementing the gate and our first step is to define a figure of merit that captures what makes a state an effective resource. These states are also fundamentally interesting, being witnesses of the difficult to achieve and highly unstable nonlinear dynamics already studied for classical mechanical systems~\cite{siler2017thermally,siler2018diffusing}.    

Ideally, the resource should have a vanishing first moment of $p_\text{NLQ}$ in order to avoid systematic displacements. Nevertheless the first moment is somewhat trivial as non-zero values can be corrected via classical displacements on the quadratures of the output state. More importantly, if the fluctuations of the nonlinear quadrature $p_\text{NLQ}$ are below the level set by the vacuum then we observe nonlinear squeezing. This means that the application of the nonlinear phase gate will have a noise performance superior to that of the ground state of a system in a quadratic potential. Operationally this guarantees that the nonlinear phase gate will work better than any classical counterpart based on classical coherent states and nonlinear adaptive feedforward control. In general, the definition of the function describing these fluctuations is
\begin{align}
V^{(n)}[\rho](\lambda)&\coloneqq\text{Var}(p_\text{NLQ})\\&\equiv\braket{(p-n\lambda q^{n-1})^2}_\rho-\braket{p-n\lambda q^{n-1}}^2_\rho \nonumber\,,\label{NLSdef}
\end{align}
Note that the ideal nonlinear phase state $e^{i\gamma_G q^n}\ket{p=0}$ has the value zero for both the mean and variance of $p_\text{NLQ}$ at $\lambda=\gamma_G$. Herein we pay attention only to the specific case of the cubic nonlinearity. The methodology presented however, is easily generalised to higher orders of nonlinear squeezing. 

The nonlinear squeezing evaluated on the vacuum is
\begin{equation}\label{NLSVac}
V[\ket{0}\bra{0}](\lambda)=\frac 12(1+9\lambda^2)\,,
\end{equation}
which is never zero and increases for larger values of the nonlinear term. In order for a state $\rho$ to qualify as a resource for implementing the cubic phase gate $e^{i\gamma_G q^3}$ that is superior to using the ground state it must satisfy the property $V[\rho](\gamma_G)<V[\ket{0}\bra{0}](\gamma_G)$ i.e. the resource performs better than the ground state in performing the gate. In general there will be a range of values of $\lambda$ over which the resource surpasses the vacuum in quality. The greater the nonlinear squeezing, the greater the value of the resource for the measurement-induced implementations. This evaluation can be extended to compare the resource with any classical state represented by a mixture of coherent states used to implement the phase gate. As a consequence, there is a threshold for nonclassical states of the oscillator which is always surpassed by states with reduced fluctuations in $p_\text{NLQ}$.

In order to demonstrate the equivalence between detecting nonlinear squeezing and nonclassicality through the nonlinear quadrature we first observe that displacements in momentum do not change the value of $V[\rho](\lambda)$. This can be quite easily seen as follows:
\begin{align}
&\braket{(p+\bar{p}-3\lambda q^2)^2}-\braket{p+\bar{p}-3\lambda q^2}^2=\\&=\braket{(p-3\lambda q^2)^2}+\braket{\bar{p}^2+2\bar{p}(p-3\lambda q^2)}-\\&~~~~~\braket{p-3\lambda q^2}^2-\braket{\bar{p}^2+2\bar{p}(p-3\lambda q^2)}\nonumber\\
&=\braket{(p-3\lambda q^2)^2}-\braket{p-3\lambda q^2}^2\,,
\end{align}
where $\bar{p}$ denotes a displacement in momentum. It follows that for each state $\rho$ there is a displaced state $\rho_D$ whose second moment, defined similarly as $V_2[\rho](\lambda)=\braket{(p-3\lambda q^2)^2}_\rho$, has the same value as $V[\rho](\lambda)$. More simply, we have the equality
\begin{equation}
V[\rho](\lambda)=V_2[\rho_D](\lambda)\,,\label{displaceequiv}
\end{equation}
for a displacement $D$ chosen such that $\braket{p}=3\lambda\braket{q^2}$. With this in mind, we proceed to demonstrate that nonlinear squeezing implies nonclassicality of the kind captured by the Glauber-Sudarshan $P$-function, with $P(\alpha)\ge0$ i.e. $\rho_\alpha=\int d^2\alpha P(\alpha)\ket{\alpha}\bra{\alpha}$ with $d^2\alpha=d\text{Re}(\alpha)d\text{Im}(\alpha)$.

As previously defined, nonlinear squeezing occurs whenever $V[\rho](\lambda)<V[\ket{0}\bra{0}](\lambda)$, for any given $\lambda\in\mathbb{R}$, and the threshold for cubic nonlinear squeezing is $V[\ket{0}\bra{0}](\lambda)=\frac 12(1+9\lambda^2)$.  Furthermore, we note that since displacements do not produce nonclassicality, we can displace any state without the risk of evaluating a classical state as nonclassical. More specifically, we can perform the displacement described in Eq.~\ref{displaceequiv}. That is, if we find a lower bound for the second moment, we also find a lower bound for the variance. 

For coherent states then, the lower bound on $V_2$ is given by
\begin{equation}
V_2[\ket{\beta}\bra{\beta}](\lambda)=\frac 12(1+9\lambda^2)\,,
\end{equation}
where $\ket{\beta=\frac{3i\lambda}{2\sqrt{2}}}$ is explicitly dependent on $\lambda$. This is sufficient to show that any mixture of coherent states is also bounded by this quantity. Consider the following inequalities:
\begin{align}
\frac 12(1+9\lambda^2)&\le\braket{\alpha|p^2_\text{NLQ}|\alpha}\\
\Rightarrow\frac 12(1+9\lambda^2)\int d^2\alpha P(\alpha)&\le\int d^2\alpha P(\alpha)\braket{\alpha|p^2_\text{NLQ}|\alpha}\\
\Rightarrow\frac 12(1+9\lambda^2)&\le\text{tr}(p_\text{NLQ}^2\rho_\alpha)\,.
\end{align}
Thus the threshold for nonclassicality is identical to that for nonlinear squeezing. It follows that a state that displays nonlinear squeezing must also be nonclassical. The converse is similar; if a state shows nonclassicality in the variance of $p_\text{NLQ}$, then it also has nonlinear squeezing. Nonlinear squeezing therefore shares this equivalence with linear squeezed states~\cite{loudon1987squeezed,lvovsky2015squeezed}. We reiterate however that nonlinear squeezing may occur independently of traditional linear squeezing.

\subsection{Direct Detection Method}

Mechanical systems capable of being influenced by a nonlinear potential are probed by an optical beam and therefore any estimation of nonlinear squeezing is indirect and influenced by this coupling and the associated optical noise. To include a broad class of experimental realisations we consider optomechanical systems whose Hamiltonian dynamics is characterised by that of a standard model~\cite{aspelmeyer2014cavity} in which a mechanical oscillator is driven by an external laser field of frequency $\omega_L$ and the cavity dissipates at a rate $\kappa$. This description also covers prospective levitated optomechanical systems in cavities~\cite{chang2009cavity,kiesel2014cavity,fonseca2016nonlinear} and in particular such setups allow us to enhance the gain of the measurement through pulsed schemes with high-Q cavities. Such systems are typically described, in units of $\hbar$ and after a suitable linearisation, with the Hamiltonian
\begin{equation}
H=\Delta a^\dagger a+\Omega b^\dagger b+g(a+a^\dagger)(b+b^\dagger)\,,
\end{equation}
where $a$ and $b$ are, respectively, the cavity and mechanical annihilation operators, $\Delta=\omega-\omega_L$ is the detuning of the cavity and $\omega$ is the cavity resonance frequency, $\Omega$ is the mechanical frequency and $g$ is the interaction strength enhanced by the intensity of the laser field. 

\begin{figure}[h]
\includegraphics[width=\columnwidth]{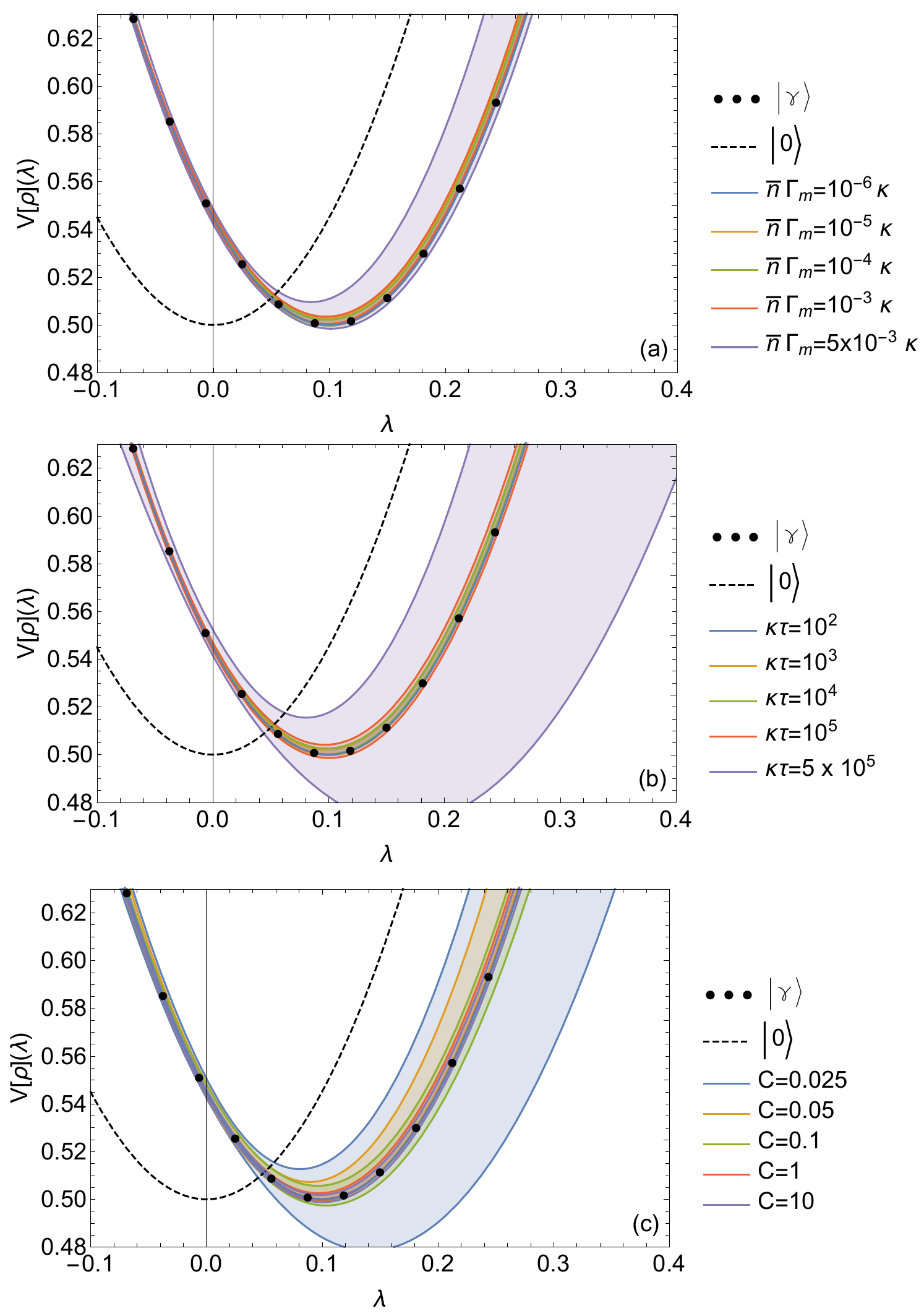}
\caption{The quality of the reconstruction procedure for the nonlinear squeezing (NLS) of the state $\ket{\gamma}$ [see Eq.~(\ref{approxCPS})] as a function of (a) the thermalisation rate $\bar{n}\Gamma_m$, (b) the interaction time $\tau$ and (c) the cooperativity $C$. The system parameters are, in dimensionless units, in (a) $G=0.1\kappa$ and $\tau\kappa=10^{3}$, in (b) $G=0.1\kappa$, $\bar{n}\Gamma_m=10^{-5}\kappa$ and in (c) $\bar{n}\Gamma_m=10^{-4}\kappa$ and $\kappa\tau=10^{3}$. The black dots trace the analytic NLS for $\ket{\gamma}$ [see Eq~(\ref{gammaNLS})] with nonlinearity $\gamma=0.1$. The dashed curve represents $V[\rho](\lambda)$ for the ground state \textit{as well as the threshold for nonclassicality}. The shaded regions indicate the error in the reconstruction for the varied parameter i.e. the upper and lower bounds to a given shaded region show the error as the standard deviation from the average reconstructed nonlinear squeezing curve. These statistical quantities are calculated from an ensemble of $20$ reconstructions, each having $10^6$ measurement results per reconstructed quadrature.}\label{Main}
\end{figure}

If one chooses to drive on resonance with the drive amplitude modulated by the mechanical frequency or, equivalently~\cite{braginsky1980quantum,aspelmeyer2014cavity}, drive with two tones on the mechanical sidebands, one achieves a QND coupling of the cavity amplitude (position) quadrature with an arbitrary mechanical quadrature $Q_\phi=\frac{be^{-i\phi}+b^\dagger e^{i\phi}}{\sqrt{2}}$~\cite{clerk2008back}. Note that we will refer to $\phi=0,\frac{\pi}{2}$ as $q$ and $p$ respectively. The phase $\phi$ is determined by the phase of the external drive.  In the frame rotating with the free mechanical energy one may invoke the rotating wave approximation to obtain the simple expression
\begin{equation}
H=GXQ_\phi\,,
\end{equation}
where $G=2g$ and $X=\frac{a+a^\dagger}{\sqrt{2}}$ is the amplitude quadrature of the cavity. The necessary condition for the RWA to hold is the resolved sideband condition $\kappa \ll \Omega$. Our choice of the QND coupling for the task of the quantum state analysis is dictated particularly by the ability of the former to perform well at moderate cavity escape efficiencies. Other more demanding options would include, for example, swapping the mechanical state to optics with subsequent optical tomography, or detection of the mechanical characteristic function via coupling to an atom (see  Ref.~\cite{vanner2015towards} and references therein). The QND interaction has been implemented in the domains of electromechanics~\cite{suh2014mechanically,wollman2015quantum,lei2016quantum} and optomechanics~\cite{shomroni2018optical}.

Alternatively, to achieve this coupling, one may consider a stroboscopic scheme on resonance in the bad cavity limit \cite{braginsky1978optimal,vanner2011pulsed}. These theoretical results on QND couplings are also available without the presence of a cavity (e.g. in levitated optomechanical setups without cavities \cite{gieseler2013thermal,rashid2016experimental}) however outside a cavity the gain of the coupling is low and this presents an efficiency that is too low to accurately reconstruct the nonlinear squeezing. Therefore our results are most relevant for the case in which the optomechanical coupling occurs within a cavity.

However the system we are examining is not unitary due to dissipation into the output signal and mechanical decoherence of the oscillator. We may write the Langevin equations~\cite{gardiner2000quantum} for the system using the input-output formalism,
\begin{align}
\dot{a}&=\frac{-iGQ_\phi}{\sqrt{2}}-\frac{\kappa}{2}a+\sqrt{\kappa}a_\text{in}\\
\dot{b}&=\frac{-iGe^{i\phi}X}{\sqrt{2}}-\frac{\Gamma_m}{2}b+\sqrt{\Gamma_m}\xi\,,
\end{align}
where $\Gamma_m$ is the mechanical damping, $a_\text{in}$ is the input cavity field and the quantum Langevin force $\xi$ describes the mechanical decoherence. This thermal noise has the following statistical properties:
\begin{align}
\braket{\xi^\dagger(t)\xi(t^\prime)}&=\bar n\delta(t-t^\prime)\,,\\
\braket{\xi(t)\xi^\dagger(t^\prime)}&=\left(\bar n+1\right)\delta(t-t^\prime)\,,
\end{align}
where $\bar n$ is the mean phonon occupation of the thermal bath. To consider the output cavity mode described by $a_\text{out}$ we make use of the input-output relation~\cite{gardiner1985input}
\begin{equation}
a_\text{in}+a_\text{out}=\sqrt{\kappa}a\,.
\end{equation}

In studying the mechanical thermal noise we note that the thermal states obey Gaussian statistics and so we may write the higher even moments $\mathcal{E}_n$ in terms of the second moment (since the first is zero). Define
\begin{equation}
\mathcal{E}_k=\braket{\mathbb{E}^k}=\begin{cases}
0 & k \text{ odd}\\
\left(\bar n+\frac 12\right)^{\frac k2}(k-1)!!\,& k \text{ even}
\end{cases}
\end{equation}
The statistics of the evolved output cavity (phase) momentum quadrature $\mathbb{Y}_\text{out}(\tau)=\frac{i(a_\text{out}^\dagger-a_\text{out})}{\sqrt{2}}$ are given by (see Appendix for greater detail)
\begin{multline}
\braket{\mathbb{Y}_\text{out}(\tau)^n}=\sum_{k_1+k_2+k_3=n}\binom{n}{k_1,k_2,k_3}\mathcal{V}_{k_1}\\\times\left(-2G\sqrt{\frac{2\tau}{\kappa}}+G\sqrt{\frac{2\tau^3}{\kappa}}\Gamma_m\right)^{k_2}\braket{Q_\phi^{k_2}}\\\times\left(-2G\tau\sqrt{\frac{2\Gamma_m}{3\kappa}}\right)^{k_3}\mathcal{E}_{k_3}\,.\label{DecReconApprox}
\end{multline}
where $\tau$ is the interval over which the interaction takes place,
\begin{align}
\mathcal{V}_k=\begin{cases}
0 & k \text{ odd}\\
\frac{1}{\sqrt\pi}\Gamma\left(\tfrac{k+1}{2}\right)& k \text{ even}
\end{cases}
\end{align}
and $\Gamma$ is the Gamma function.

Since we make no assumptions on the form of the mechanical state, Eq.~(\ref{DecReconApprox}) clearly gives us the tools to retrieve the necessary moments for mechanical $q$ and $p$ in order to construct $V[\rho](\lambda)$ and $V_2[\rho](\lambda)$. To clarify further, fixing a particular quadrature for reconstruction fixes a phase $\phi$ of the laser drive. By adjusting $\phi$ (and therefore $Q_\phi$) information about different quadratures can be copied into the momentum quadrature of the output field. However the mixed moments $\braket{pq^2}$ and $\braket{q^2p}$ are not directly available. Fortunately, we do not need full state tomography to specify them. Instead, these may be obtained by considering the rotated mechanical quadratures $Q_\frac{\pi}{4}$ and $Q_{-\frac{\pi}{4}}$, obtained by selecting appropriate phases of the external drive.  Then,
\begin{equation}
pq^2=\frac{\sqrt{2}}{3}(Q_\frac{\pi}{4}^3-Q_{-\frac{\pi}{4}}^3)-\frac{p^3}{3}-iq\,.\label{mixedmode1}
\end{equation}
Similarly,
\begin{equation}
q^2p=\frac{\sqrt{2}}{3}(Q_\frac{\pi}{4}^3-Q_{-\frac{\pi}{4}}^3)-\frac{p^3}{3}+iq\,,\label{mixedmode2}
\end{equation}
by taking advantage of commutation relations.
In summary, to construct the nonlinear squeezing function, we require moments of $q^n$ with $n=1,2,4$, $p^n$ with $n=1,2,3$ and $Q_{\pm\frac{\pi}{4}}^3$. That said, due to the hierarchical nature of Eq.~(\ref{DecReconApprox}) we must also reconstruct the first order moments of the rotated quadratures in order to retrieve the third order moments. We make a small aside here to note that higher orders of nonlinear squeezing are also accessible to this scheme, since moments up to order $n$ can be accessed by Eq.~(\ref{DecReconApprox}). The one seeming complication to this is the presence of higher order mixed moments, $\braket{pq^n}$ and $\braket{q^np}$, which can be obviated in the same manner as presented above.

Typically the parameters of an optomechanics setup, particularly those considered here such as the cavity decay and mechanical decoherence rates, and the optomechanical coupling strength, are well-characterised and stable \cite{gorodetksy2010determination,purdy2015optomechanical,ang2013optomechanical}. Given that this is so, we are in a position to evaluate the quality of a reconstruction based on the information retrievable using these relations. Consider having access to many copies of a given quantum state. By engineering the QND interaction with an appropriate $Q_\phi$ one may sample from $\mathbb{Y}_\text{out}(\tau)$ by performing homodyne detection on the output cavity field after an interaction time $\tau$.  This generates a histogram from which one may estimate the various moments of $\mathbb{Y}_\text{out}(\tau)$. Inverting the equations generated by Eq.~(\ref{DecReconApprox}) produces the moments of the chosen mechanical quadrature $Q_\phi$~\cite{wieczorek2015optimal,moore2016quantum}. Given the correct assortment of reconstructed statistics one may construct the functions $V[\rho](\lambda)$ and $V_2[\rho](\lambda)$ for the mechanical state without the necessity of performing full tomography of the mechanical state. 

\subsection{Nonlinear Squeezing}

Nonlinear squeezing can be generated by a nonlinear potential $V(q)=\gamma q^3$ temporarily influencing the mechanical oscillator while in the ground state, as depicted in Fig.~\ref{OMsketch}. If this application is sufficiently fast and with a strong enough nonlinearity, we can expect an approximate cubic phase state of the mechanical oscillator:
\begin{equation}
\ket{\gamma}=e^{i\gamma q^3}\ket{0}\,\label{approxCPS}
\end{equation}
where $\ket{0}$ is the ground state. The design of such pure states in optomechanics has been approached in the literature already~\cite{houhou2018unconditional}. Fig.~\ref{Main} demonstrates the quality of the reconstruction
under certain relevant experimental conditions (see figure caption). The figure demonstrates that the reconstruction is quite robust to mechanical decoherence over a wide range of parameters, mainly due to a short interaction time $\tau$ and mechanical decoherence characterised by rethermalisation rates $\bar{n}\Gamma_m<\frac{1}{\tau}$. Indeed, within the acceptable parameter ranges specified the reconstruction shows little bias with respect to over- or underestimating the nonlinear squeezing and low error due to statistical fluctuations in the reconstruction. The dashed line shows the fluctuations of $p_\text{NLQ}$ for the ground state which also functions as a bound for nonclassicality while the dotted curve shows the ideal nonlinear squeezing for $\ket{\gamma}$ with $\gamma=0.1$, a conservative value. The nonlinear squeezing, for various parameters, is constructed out of a parabola in $\lambda$ whose coefficients are the statistical quantities retrieved by measurement of the output cavity field. The boundaries of the shaded regions denote the error (one standard deviation) in the reconstruction of the nonlinear squeezing. That is, an ensemble of such curves was reconstructed and the error is calculated over this ensemble. The reconstruction quality experiences a sharp decrease in accuracy after passing certain thresholds in the parameters directly related to mechanical decoherence and the quality of the cavity-mechanical coupling. 

We divide the parameters into two major classes: rethermalisation, involving $\bar{n}$, $\Gamma_m$ and $\tau$, and cooperativity, mainly involving $G$ and $\kappa$. Panels (a) and (b) indicate the effects of surpassing the rethermalisation time $\Gamma_m\bar{n}\tau\ll1$. Once this threshold is crossed errors accumulate and the variance in the reconstruction of the nonlinear squeezing curve becomes very large. This can be understood taking the view that information about the mechanical state must be extracted faster than the rethermalisation time. Additionally in panel (c) we take a limiting case of the rethermalisation time and investigate the effect of changing the ratio between the coupling strength and the cavity dissipation rate in terms of the cooperativity 
\begin{equation}
C=\frac{G^2}{\bar{n}\Gamma_m\kappa}\,.
\end{equation}
The results show that the quality of the reconstruction is maintained for cooperativity values of $C\gtrsim0.1$. Advantageously, a cooperativity of $C>1$ is not required. The QND interaction required for the reconstruction is already available in electromechanics setups~\cite{wollman2015quantum,pirkkalainen2015squeezing}, where dissipative engineering may soon be capable of providing nonlinear states. Furthermore, experiments in levitated optomechanics are beginning to comfortably reach this regime~\cite{gieseler2012subkelvin,jain2016direct,vovrosh2017parametric} and have also demonstrated the capacity for applying nonlinear external potentials to a mechanical oscillator~\cite{siler2018diffusing}.

It is important that the reconstruction is accurate as any assessment of the quality of the resource derived from the reconstruction will be benchmarked against the ground state. The errors in the reconstruction must not be so wide that the error curves (one standard deviation) everywhere cross the nonclassicality benchmark so that the resource cannot be distinguished from classical resources. Our results indicate favourably that the largest error in the reconstruction occurs for values of $\lambda>0.1$ for which the nonlinear squeezing is far from the threshold set by the ground state. Additionally, one must take care to have minimal error in order to prevent an overestimation of the quality of the resource. For example, once the relevant thresholds are surpassed it is possible, in the worst case, to greatly overestimate the value of the nonlinear squeezing. In this case it may be reasonable to use the upper limits as conservative estimates. On the other hand, even a weak nonlinearity is sufficient to surpass the ground state limit and since the ground state also represents the bound for nonclassicality any nonlinearly squeezed state is inherently nonclassical.

To illustrate the effectiveness of searching for nonlinear squeezing we again assume the approximate cubic phase state $\ket{\gamma}$ and show how it provides an advantage over the vacuum state for a range of values of $\gamma$. The nonlinear squeezing for this state is
\begin{equation}
V[\ket{\gamma}\bra{\gamma}](\lambda)=\frac 12(1+9(\gamma-\lambda)^2)\,.\label{gammaNLS}
\end{equation}
Recall that in order for $\ket{\gamma}$ to constitute a resource for applying the gate $e^{i\gamma_Gq^3}$ the state must satisfy the condition $V[\rho](\gamma_G)<V[\ket{0}\bra{0}](\gamma_G)$. Assuming $\gamma>0$ it is clear that this occurs whenever $\gamma<2\gamma_G$.

Naturally, the approximate cubic phase state is most effective as a resource whenever $\gamma=\gamma_G$. However, exact matching of the nonlinearity of the cubic resource state and that of the cubic phase gate is not necessary in order to gain an advantage on classical states. As said, in practice one may not know in advance what state has been prepared. We stress that the method of reconstructing the nonlinear squeezing provides an opportunity to extract the quality of the resource via measurements on a few quadratures of light, requiring significantly less effort than full tomography. 

\section{Discussion}

The main result of this article is the provision of a method for reconstruction of the fluctuations in a nonlinear combination of quadratures of a mechanical mode in optomechanics without complete mechanical tomography. In particular we focus on the nonlinear quadrature generated in momentum by a cubic potential, relevant for noise reduction in nonlinear circuits employing the celebrated cubic phase gate. We provide an analysis of the robustness of this reconstruction method in the context of the cooperativity and mechanical decoherence. This is important for nonlinear states displaying reduced fluctuations in such a quadrature as the properties emerging from the nonlinearity are susceptible to being wiped out by the Gaussian noise of a thermal bath. States which exhibit nonlinear squeezing in this regard are also shown to exhibit $P$-function nonclassicality similar to Gaussian squeezed states from linearised dynamics. To the advantage of state of the art experiments, even weak nonlinearities display significant nonlinear squeezing compared to the ground state. It is straightforward to extend this methodology to higher orders of nonlinear potentials to detect the aspects relevant for the construction of nonlinear phase gates~\cite{marek2018general}. 


The setting presented here is very general for opto- and electromechanics but we would like to emphasise the applicability of our scheme to levitated systems~\cite{gieseler2012subkelvin,jain2016direct,vovrosh2017parametric}, given the large range of decoherence parameters over which the scheme is viable. As mentioned, levitated systems are able to employ nonlinear potentials for the dynamics of the levitated particle~\cite{siler2018diffusing}, thus enabling the preparation of nonlinear states, and have already approached the regimes in which the rethermalisation time threshold can be met. The major challenge for the future is to achieve QND couplings with the levitated system and incorporate nonlinear state preparation into a single setup. 

\section{Acknowledgements}


The authors have received national funding from the MEYS of the Czech Republic (Project No. 8C18003) under Grant Agreement No. 731473 within the QUANTERA ERA-NET Cofund in Quantum Technologies implemented within the European Union's Horizon 2020 Programme (ProjectTheBlinQC). We also acknowledge support from the Czech Science Foundation under project 19-17765S. DM acknowledges support from the Development Project of Faculty of Science, Palack\'{y} University. AR acknowledges project LTC17086 of the INTER-EXCELLENCE program of the Czech Ministry of Education.

%
%
%

\bibliography{references}

\widetext
\clearpage

\setcounter{page}{1}%
\setcounter{figure}{0}%
\setcounter{equation}{0}%
\setcounter{section}{0}
\setcounter{table}{0}
\renewcommand{\theequation}{S\arabic{equation}}
\renewcommand{\thefigure}{S\arabic{figure}}
\renewcommand{\thepage}{S\arabic{page}}
\renewcommand{\thesection}{S\arabic{section}}
\renewcommand{\thetable}{S\Roman{table}}

\appendix

\section{Hamiltonian Derivation}

Here we provide a short derivation of the Hamiltonian in Eq.~(12). We begin, as in the text, with a suitably linearised optomechanical Hamiltonian
\begin{equation}
H=\Delta a^\dagger a+\Omega b^\dagger b+g(a+a^\dagger)(b+b^\dagger)\,.
\end{equation}
In the main text we refer to a two tone driving involving drive tones on both red and blue sidebands. The references provided give greater detail on how the QND interaction can be derived from such a scenario. Essentially, the two tone drive involves a fast oscillation on resonance ($\Delta=0$) with the cavity frequency and a slower envelope oscillating at the mechanical frequency. Here we provide a simplified explanation assuming on resonance driving and an interaction profile oscillating with the mechanical frequency. In the frame rotating with the mechanical frequency this corresponds to the Hamiltonian
\begin{equation}
H=G\cos(\Omega t+\phi)X(be^{-i\Omega t}+b^\dagger e^{i\Omega t})\,,
\end{equation}
where $\phi$ is an arbitrary phase of the external drive and we have rewritten the cavity mode in terms of the position (amplitude) quadrature $X=\frac{a+a^\dagger}{\sqrt{2}}$. Now with some algebra one may rewrite the Hamiltonian as
\begin{equation}
H=GX(Q_\phi+q\sin\Omega t+p\cos\Omega t)\,,
\end{equation}
with $Q_\phi=\frac{be^{-i\phi}+b^\dagger e^{i\phi}}{\sqrt{2}}$, $Q_0=q$ and $Q_{\frac{\pi}{2}}=p$. Assuming the rotating wave approximation allows us to drop the time dependent terms and what results is Eq.~(12).

\section{Input-Output Theory}

Here we give a fuller account of the derivation of Eq.~(\ref{DecReconApprox}) in the main text through the apparatus of input-output theory. This also includes the assumptions and approximations we have made on the system dynamics. The system we are examining is not unitary due to dissipation on the cavity field and mechanical decoherence on the resonator. We may write the Langevin equations for the system using the input-output formalism,
\begin{align}
\dot{a}&=\frac{-iGQ_\phi}{\sqrt{2}}-\frac{\kappa}{2}a+\sqrt{\kappa}a_\text{in}\\
\dot{b}&=\frac{-iGe^{i\phi}X}{\sqrt{2}}-\frac{\Gamma_m}{2}b+\sqrt{\Gamma_m}\xi\,,
\end{align}
where $\Gamma_m$ is the mechanical damping, $a_\text{in}$ is the input cavity field and $\xi$ describes the mechanical decoherence.

Constructing the quadratures from these equations,
\begin{align}
\dot{X}&=-\tfrac{\kappa}{2}X+\sqrt{\kappa}X_\text{in}\\
\dot{Y}&=-\sqrt{2}GQ_\phi-\tfrac{\kappa}{2}Y+\sqrt{\kappa}Y_\text{in}\\
\dot{Q}_\phi&=-\tfrac{\Gamma_m}{2}Q_\phi+\sqrt{\tfrac{\Gamma_m}{2}}(\xi e^{-i\phi}+\xi^\dagger e^{i\phi})\label{inoutQ}\,,
\end{align}
one may require that $\kappa$ is the dominant frequency, which allows the cavity field to adiabatically follow the dynamics. In this case the cavity momentum is simply expressed as
\begin{equation}
Y=\tfrac{2}{\kappa}(\sqrt{\kappa}Y_\text{in}-\sqrt{2}GQ_\phi)\,.
\end{equation}
To consider the output cavity momentum we make use of the input-output relation
\begin{equation}
a_\text{in}+a_\text{out}=\sqrt{\kappa}a\,.
\end{equation}
Then,
\begin{equation}
Y_\text{out}(t)=Y_\text{in}(t)-2G\sqrt{\tfrac{2}{\kappa}}Q_\phi(t)\,.
\end{equation}
If mechanical decoherence is neglected, $Q_\phi(t)\equiv Q_\phi^0$.
The homodyne detector measures a certain temporal mode of the leaking field defined by
\begin{align}
\mathbb{Y}_\text{out}(\tau)=\int_0^\tau Y_\text{out}(t)f_\text{out}(t)dt
~~~~\Rightarrow~~~~\mathbb{Y}_\text{out}=\mathbb{Y}_\text{in}(\tau)-2G\sqrt{\tfrac{2}{\kappa}}Q_\phi^0\int_0^\tau f_\text{out}(t)dt\,.
\end{align}
If $f_\text{out}=\frac{1}{\sqrt{\tau}}$ then we simply have that
\begin{equation}
\mathbb{Y}_\text{out}(\tau)=\mathbb{Y}_\text{in}(\tau)-2G\sqrt{\tfrac{2\tau}{\kappa}}Q_\phi^0\,.
\end{equation}
Then it follows that the statistics of the output momentum are represented by
\begin{equation}\label{InOutRecon}
\braket{\mathbb{Y}_\text{out}(\tau)^n}=\sum_k\binom{n}{k}\mathcal{V}_k\left(-2G\sqrt{\tfrac{2\tau}{\kappa}}\right)^{n-k}\braket{Q_\phi^{n-k}}\,.
\end{equation}
At time $\tau$ the interaction is held to have been switched off hence the input field is in the vacuum and has statistics represented by $\mathcal{V}$. It is clear that the process of inverting this hierarchy of equations depends on the interplay between the set of parameters $\{G,\tau,\kappa\}$.

What remains is to develop the effect of thermal decoherence on the reconstruction procedure. The thermal noise $\xi$ introduced above has the following statistical properties:
\begin{align}
\braket{\xi^\dagger(t)\xi(t^\prime)}=\bar n\delta(t-t^\prime)\,,
~~~~\braket{\xi(t)\xi^\dagger(t^\prime)}=\left(\bar n+1\right)\delta(t-t^\prime)\,,
\end{align}
where $\bar n$ is the average occupation of the bath. In this case we must examine Eq.~(\ref{inoutQ}) in the context of nonzero $\Gamma_m$. The formal solution to this equation has the form
\begin{equation}
Q_\phi(t)=Q_\phi(0)e^{-\frac{\Gamma_mt}{2}}+\sqrt{\Gamma_m}e^{-\frac{\Gamma_mt}{2}}\int_0^t\xi_\phi(s)e^{\frac{\Gamma_ms}{2}}ds\,,
\end{equation}
where $\xi_\phi=\frac{\xi e^{-i\phi}+\xi^\dagger e^{i\phi}}{\sqrt{2}}$. Applying the rectangular mode filter $f_\text{out}=\frac{1}{\sqrt{\tau}}$ to this equation results in 
\begin{equation}
\mathbb{Q}_\phi(\tau)=\frac{2Q_\phi(0)}{\Gamma_m\sqrt{\tau}}(1-e^{-\frac{\Gamma_m\tau}{2}})-2\sqrt{\frac{\tau\Gamma_m-3+4e^{-\frac{\Gamma_m\tau}{2}}-e^{\Gamma_m\tau}}{\Gamma_m^2\tau}}\mathbb{E}_\phi\,.
\end{equation}
Note that $\mathbb{E}_\phi$ is a proper quadrature of the field (obeying canonical commutation relations) defined by 
\begin{align}
\mathbb{E}_\phi=\frac{\int_0^\tau\xi_\phi(e^{\frac{\Gamma_m}{2}(s-\tau)}-1)ds}{\sqrt{\int_0^\tau(e^{\frac{\Gamma_m}{2}(s-\tau)}-1)^2ds}}=\sqrt{\frac{\tau\Gamma_m-3+4e^{-\frac{\Gamma_m\tau}{2}}-e^{\Gamma_m\tau}}{\Gamma_m}}\int_0^\tau\xi_\phi(e^{\frac{\Gamma_m}{2}(s-\tau)}-1)ds\,.\nonumber
\end{align}
Finally, the relation between the output cavity mode and the mechanical quadratures is given by
\begin{equation}
\mathbb{Y}_\text{out}(\tau)=\mathbb{Y}_\text{in}(\tau)-\frac{4GQ_\phi(0)}{\Gamma_m}\sqrt{\frac{2}{\kappa\tau}}(1-e^{-\frac{\Gamma_m\tau}{2}})-4G\sqrt{\frac{2(\Gamma_m\tau+4e^{-\frac{\Gamma_m\tau}{2}}-e^{\Gamma_m\tau}-3)}{\kappa\tau\Gamma_m^2}}\mathbb{E}_\phi\,.\label{MechDecOp}
\end{equation}
In the limit $\Gamma_m\rightarrow0$ this expression recovers what has already been derived. Following from this, the statistics of the output field are related to the mechanical quadrature moments via
\begin{multline}
\braket{\mathbb{Y}_\text{out}(\tau)^n}=\sum_{k_1+k_2+k_3=n}\binom{n}{k_1,k_2,k_3}\mathcal{V}_{k_1}\left(-\frac{4G}{\Gamma_m}\sqrt{\frac{2}{\kappa\tau}}(1-e^{-\frac{\Gamma_m\tau}{2}})\right)^{k_2}\braket{Q_\phi(0)^{k_2}}\\\times\left(-4G\sqrt{\frac{2(\Gamma_m\tau+4e^{-\frac{\Gamma_m\tau}{2}}-e^{-\Gamma_m\tau}-3)}{\kappa\tau\Gamma_m^2}}\right)^{k_3}\braket{\mathbb{E}_\phi^{k_3}}\,.\label{DecRecon}
\end{multline}

It is straightforward to show that $\mathbb{E}_\phi$ obeys Gaussian statistics over the thermal state and furthermore is symmetric under rotations in phase space i.e. we may omit the angle $\phi$. The higher even moments of $\mathbb{E}$ are given in terms of the second moment (since the first is zero). Define
\begin{equation}
\mathcal{E}_n=\braket{\mathbb{E}^n}=\begin{cases}
0 & n \text{ odd}\\
\left(\bar n+\frac 12\right)^{\frac n2}(n-1)!!\,& n \text{ even}.
\end{cases}
\end{equation}

A further simplification can be readily achieved by expanding the coefficients to first order in $\Gamma_m$. The result is Eq.~(\ref{DecReconApprox}) in the main text.

\end{document}